\documentclass[aps,amssymb]{revtex4}
\usepackage{graphicx,psfrag}

\renewcommand{\a}{\alpha}

\newcommand{\G}{\Gamma}
\renewcommand{\d}{\delta}\newcommand{\D}{\Delta}
\newcommand{\e}{\epsilon}

\renewcommand{\l}{\lambda}\renewcommand{\L}{\Lambda}
\renewcommand{\t}{\theta}\newcommand{\T}{\Theta}

\newcommand{\s}{\sigma}
\newcommand{\p}{\phi}\renewcommand{\P}{\Phi}
\renewcommand{\r}{\rho}
\renewcommand{\u}{\mu}
\renewcommand{\v}{\nu}
\renewcommand{\o}{\omega}

\newcommand{\vj}{\vec{\jmath}}
\newcommand{\vm}{\vec{m}}

\newcommand{\cA}{{\mathcal A}}

\newcommand{\cC}{{\mathcal C}}

\newcommand{\cN}{{\mathcal N}}

\newcommand{\cT}{{\mathcal T}}

\newcommand{\real}{\mathbb{R}}
\newcommand{\comp}{\mathbb{C}}

\newcommand{\SO}{\mathrm{SO}}
\newcommand{\SU}{\mathrm{SU}}

\newcommand{\su}{\mathrm{su}}

\renewcommand{\dim}{\mathrm{dim}}

\newcommand{\R}{\mathfrak{R}}

\newtheorem{theo}{Theorem}

\newcommand{\be}{\begin{equation}}
\newcommand{\ee}{\end{equation}}
\newcommand{\beq}{\begin{eqnarray}}
\newcommand{\eeq}{\end{eqnarray}}

\renewcommand{\v}[1]{\vec{#1}}
\newcommand{\tj}[2]{C^{\vec{#1}}_{\vec{#2}}}

\newcommand{\sj}[1]{\left\{\begin{array}{ccc}
{#1}_1 & {#1}_2 & {#1}_3 \\
{#1}_4 & {#1}_5 & {#1}_6
\end{array}\right\}}

\newcommand{\tZ}{\tilde{Z}}

\begin{document}
\begin{titlepage}
\title{\large Non-perturbative summation over 3D discrete topologies}
\author{ Laurent Freidel}
\email{lfreidel@perimeterinstitute.ca}
\affiliation{\vspace{2mm}Perimeter Institute for Theoretical
Physics\\ 35 King street North, Waterloo  N2J-2G9,Ontario,
Canada\\} \affiliation{Laboratoire de Physique, \'Ecole Normale
Sup{\'e}rieure de Lyon \\ 46 all{\'e}e d'Italie, 69364 Lyon Cedex
07, France }

\author{David Louapre}
\email{dlouapre@ens-lyon.fr} \affiliation{\vspace{2mm}Laboratoire
de Physique, \'Ecole Normale Sup{\'e}rieure de Lyon \\ 46
all{\'e}e d'Italie, 69364 Lyon Cedex 07, France}\thanks{UMR 5672
du CNRS}

\date{\today}

\begin{abstract}
\begin{center}{\bf Abstract }\end{center}
We construct a group field theory which realizes the sum of
gravity amplitudes over all three dimensional topologies  trough a
perturbative expansion. We prove this theory to be uniquely Borel
summable. This shows how to define a non-perturbative summation
over triangulations including all topologies in the context of
three dimensional discrete gravity.
\end{abstract}
\maketitle
\end{titlepage}

\section{Introduction}
In quantum gravity, we are interested in the computation of
transition amplitudes between spatial geometries. These transition
amplitudes are supposed to arise as a sum over spacetime
`geometries'. In the classical realm what is meant by geometries
is clear, one usually means a smooth Lorentzian metric free of
singularity on a manifold of fixed topology, possibly satisfying
some asymptotic conditions. In the quantum realm we obviously have
much more freedom in choosing which spacetime geometries should be
included in the sum. Do we authorize the geometries to posses some
singularities for instance? \cite{Horowitz} Do we authorize topology change? One
of the outstanding question in this context is wether one should
or not include a sum over topology in the sum over geometries. The
next outstanding question is wether this is possible in a given
framework. In this paper we are going to analyze this problem in
the context of discrete approach to quantum gravity, i-e dynamical
triangulation and spin foam models. In dynamical triangulation one
first choose an equilateral triangulation $\Delta$ of space-time
with regular simplices and we assign to each triangulation a
statistical weight given by the exponential of the Regge action.
In spin foam models one choose a triangulation of space time and
assign to it certain representation labels of the Lorentz group.
This representation labels carry information on the geometry of
the simplex. The transition amplitudes are constructed as a state
sum, using sum over representations assigned to the geometrical
building blocks of the triangulation. Historically, the first spin
foam model was proposed by Ponzano and Regge \cite{ponzano68} for
3D Euclidian quantum gravity, and their work has been recently
generalized to higher dimension to give birth to the spin foam
approach for quantum gravity (see \cite{baez99,OritiRev} for an
overview).

3D gravity is a topological theory. As a consequence, the
transition amplitudes built in the Ponzano-Regge model do not
depend on the choice of the triangulation. However this is no
longer the case in 4D, and one has somehow to get rid of the
choice of the triangulation in order to restore the infinite
number of degrees of freedom of 4D gravity. A proposal to do this
is to sum the discretized amplitude over all possible
triangulations of the manifold. A more drastic proposal is to sum
over all possible triangulations regardless of the topology.

It is however commonly believed that there is no way to give sense
to the sum over triangulations including all topologies. The
problem appears clearly in the context of dynamical triangulations
\cite{ambjorn91}. In this context the amplitude of a given
3-dimensional triangulation $\D$ is given by \be \cA[\D] = {1\over
\textrm{Sym}[\D]} \l^{N_3}\mu^{N_1},\label{eqn:amplDT} \ee where
$\textrm{Sym}[\D]$ is a symmetry factor, and  $N_3$ (resp. $N_1$)
denote the number of tetrahedra (resp. edges) of the
triangulation. The parameters $\lambda, \mu$ can be interpreted,
if one uses the
 Regge action,
in terms of the bare cosmological constant $\L $ and the bare
Newton constant $1/G$ \beq
\mu \sim \exp({1\over G}), \\
\lambda \sim \exp(-\L -{c\over G}), \label{eqn:setlambda}
 \eeq
 where $c$ is a geometrical constant. The
sum over all triangulations is then \be Z=
\sum_{N_3,N_1}\l^{N_3}\mu^{N_1} \cN({N_3},{N_1}), \ee with
$\cN({N_3},{N_1})$ being the number of triangulations with a fixed
number of tetrahedra and edges. Since $\mu \geq 1$ the partition
function is bounded from below by $\sum_{N_3}\l^{N_3}\cN({N_3})$
where $\cN({N_3})$ is the total number of triangulations with a
fixed number of tetrahedra. This number grows factorially with the
number of tetrahedra if there is no restriction on the topology of
the triangulation. Therefore a sum over triangulations does not
seem to be possible
 in general, since this
factorial growth  can not be killed by any choice of $\l,\mu$. The
solution proposed in \cite{ambjorn91} is to restrict to a fixed
topology (as the 3-sphere), in order to get back to a case where
the number of triangulations grows exponentially with the number
of 3-simplices. So the sum over simplicial manifold can be defined
in the case of a restricted topology. However, since there is no
general classification of the topologies in $D>2$, we cannot
expect to use this result to make sense of a sum over topologies
as in 2 dimensions. One conclusion that one can draw is that one
should not try to sum over topology when trying to solve quantum
gravity. This is not completely satisfactory, since first, there
are argument in favor of classical topology change
\cite{Horowitz}, but also one would expect in a theory of quantum
gravity to resolve this issue dynamically and not a priori.

There is hopefully another point of view that can be taken on this
problem. It is well known since Dyson \cite{dyson52} that physical
quantities in interacting quantum theory are non-analytic
functions of the coupling constant. This translates into the fact
that the perturbative expansion of any physical quantities in
terms of the coupling constant is a divergent series. This opens
the possibility of giving a meaning to the sum over triangulations
if we can interpret it as a perturbative expansion of a
non-perturbative quantity.

These ideas have already been used in the context of 2D gravity,
leading to the construction of the matrix models, which realize a
sum over triangulations of 2D surfaces \cite{ginsparg91}. By
considering three-indices tensors instead of matrices, Ambjorn et
al. \cite{ambjorn91} tried to apply these ideas to obtain a
dynamical triangulation model for three-dimensional quantum
gravity. Later, Boulatov and  Ooguri \cite{boulatov92,ooguri92}
have generalized these models in the form of non local field
theories over group manifold. The bottom line is that a space-time
triangulation can naturally be interpreted as a Feynman graph of a
Group Field Theory (GFT) \cite{Robpetro}. Moreover, the specific
gravity amplitude assigned to a given triangulation is exactly
given by the Feynman graph evaluation of a group field theory GFT.
This was shown in
the context of 3D dynamical triangulation \cite{ambjorn91}, 3D
Ponzano-Regge model \cite{boulatov92} and even 4D spin foam model
\cite{depietri98}. It is now understood that this is a general
feature of spin foam models \cite{Reis}.

All these results lead to the conclusion that the sum over
triangulations of a state sum amplitude is in fact a sum over
Feynman graphs of a GFT,
\begin{equation}\label{eq:D=G}
\sum_{\D} \cA_{\textrm{State sum}}(\D) = \sum_{\G}
\cA_{\textrm{GFT}}(\G).
\end{equation}
Therefore it is not a surprise that the sum over triangulations is
divergent, since it is a perturbative expansion, it should be
understood as an asymptotic series. If we take for serious the
fact that the sum over triangulations comes from a group field
theory, this opens the possibility of giving a non perturbative
meaning to the sum. This idea has been already advocated in
\cite{Reis} and to some extent in \cite{ambjorn91}.

However it is well known that in general, the asymptotic series
does not determine uniquely the resummation. One need additional
non-perturbative inputs. In interacting quantum field theory, the
large order behavior is  determined  by non-perturbative
information contained in the action and in some case \cite{zinn90}
one is able to reconstruct a non-perturbative expression from the
divergent series, in a {\it uniquely} defined manner. In such
cases, the divergent series is said to be uniquely Borel-summable.
Our main result in this paper is to prove that such a procedure can be
realized for a particular type of GFT models associated with three
dimensional gravity. If the Feynman expansion of a GFT model is
Borel-summable, this means it could be related to a non
perturbative  expression in a uniquely defined manner, giving a
well-defined meaning to the equality
\begin{equation}
\sum_{\G} \cA_{\textrm{GFT}}(\G) = Z_{NP}.
\end{equation}
 This result proves that a well-defined notion of a sum
over all triangulations of all topologies is not unreasonable, as
it could be achieved in particular cases.

In the context of dynamical triangulation the coupling constant
$\l$ of the group field theory has a physical meaning, it can be
interpreted as $- \exp(-\L )$, where $\L$ is the cosmological
constant. Since it is negative, the potential is unbounded from
below and it was expected that the corresponding GFT is not
uniquely Borel summable in the physical regime. Therefore one can
think that the theories allowing non perturbative resummation will
not be physical (at least in the context of dynamical
triangulations). In two dimension one can view this problem as the
one preventing a non perturbative definition of string theory via
matrix models.

It was therefore for us a big surprise and a relief to find out
that the modification of the Ambjorn et al. model \cite{ambjorn91}
we propose is uniquely Borel summable  in the {\it physical region
} where $\l>0$ (see sect. \ref{sec:modif}). In the context of
three dimensional Ponzano-Regge models the value of the  group
field theory coupling constant model is not restricted by physical
requirement and we expect the Borel summable GFT to have physical
meaning without restriction.

In section \ref{sec:gft} of this paper we first review the
definitions of several group field theory models related to 3
dimensional gravity like the Boulatov or Ambjorn model. We present
their Feynman expansions and show their relations with 3 dimensional
Ponzano-Regge models or dynamical triangulations. We also discuss
their convergence properties. In section
\ref{sec:nonpert} we review known facts and theorems concerning
non perturbative resummation of a perturbative series. In
section \ref{sec:modif} we propose a modification of the usual
Ambjorn or Boulatov models in order to deal with positive
potential. We show that  the
perturbative expansion of this model can be related to the dynamical
triangulation sum with physical value of the parameters. In the
last section \ref{sec:proof} we finally prove the unique Borel
summability of the model we have defined. This give us an example
of an unambiguous  way to sum over  all triangulations hence all
topologies.


\section{3D Group Field Theory models}\label{sec:gft}
In this part, we review various constructions of 3D group field
theory (GFT) models : the Boulatov model, leading to the Ponzano-Regge
amplitude \cite{boulatov92}, a model on the non-commutative
sphere, reproducing the 3D matrix model of Ambjorn et
al.\cite{ambjorn91}, and some others models with improved
convergence properties, analogous to those recently introduced for
4D spin foam models \cite{perez00,perez01}.
\subsection{The Boulatov model}
The Boulatov model is a field theory defined on the group manifold $SO(3)^3$, with some invariances
imposed on the fields. The fields are asked to be real ($\bar{\P}=\P$), symmetric
\begin{equation}
\P(g_{\sigma(1)},g_{\sigma(2)},g_{\sigma(3)}) = (-1)^{|\sigma|}
\P(g_1,g_2,g_3),
\end{equation}
(where $\sigma$ denotes a permutation of three elements and $|\sigma|$ its signature), and (right-)
$SO(3)$ invariant
\begin{equation}
\P(g_1g,g_2g,g_3g)=\P(g_1,g_2,g_3).
\end{equation}
This last requirement allows us to see the Boulatov model as a
field theory on the coset space $\SO(3)^3/\SO(3)$. The action is
non-local and is given by
\begin{eqnarray}
S[\P]&=&\frac{1}{2} \int_{G^3} dg_1dg_2dg_3\ \P(g_1,g_2,g_3)^2
\nonumber
\\ &+& \frac{\l}{4!} \int_{G^6}  dg_1dg_2dg_3dg_4dg_5dg_6\
\P(g_1,g_2,g_3)\P(g_3,g_5,g_4) \P(g_4,g_2,g_6) \P(g_6,g_5,g_1)
\label{eqn:action}
\end{eqnarray}

To compute the partition function of this model in perturbative
expansion, one has to express the action in terms of the  Fourier
coefficients of the field $\P$. In order to define a Fourier
transform on the coset space, one first considers the transform on
$\SO(3)^3$, then requires the $\SO(3)$-invariance. The
representations of $\SO{(3)}$ are labelled by integer spin $j$, and
a function on $\SO(3)$ can be expressed as a sum over  the matrix
elements  of $g$ in the spin $j$ representation $D^{j}_{mn}(g)$
(see appendix \ref{app:representations})
\begin{equation}
\P(g)=\sum_{j,m,n} d_j \P_{j}^{mn} D^{j}_{mn}(g),
\end{equation}
where $d_j =2j+1$ and $-j\leq m,n \leq j$. The Fourier coefficient
is  \be \P_{j}^{mn} = \int_G dg\  \P(g) D^{j}_{mn}({g^*}), \ee
where ${g^*}$ denotes complex conjugation. The Fourier expansion
on $\SO(3)^3$ is then
\begin{equation}
\P(g_1,g_2,g_3) = \sum_{\v{\jmath},\v{m},\v{n}}
d_{\v{\jmath}}  \P_{\v{\jmath}}^{\v{m}\v{n}} D^{\v{\jmath}}_{\v{m}\v{n}}(\v{g}),
\label{eqn:FourierSO3}
\end{equation}
where $\v{j}\equiv(j_1,j_2,j_3)$ and
$D^{\v{\jmath}}_{\v{m}\v{n}}(\v{g})\equiv
D^{j_1}_{m_1,n_1}(g_1)D^{j_2}_{m_2,n_2}(g_2)D^{j_3}_{m_3,n_3}(g_3)$.
The $\SO(3)$-invariance is implemented by group averaging over the
right diagonal action of $\SO(3)$
\begin{equation}
\P(g_1,g_2,g_3) = \int dg\ \P(g_1g,g_2g,g_3g).
\end{equation}
Expanding the RHS using decomposition (\ref{eqn:FourierSO3}), then
performing the integration on $g$ using the relation
(\ref{eqn:int3mat}), we obtain the Fourier decomposition of the
field over $\SO(3)^3/\SO(3)$
\begin{equation}\label{eqn:fourierdecomp}
\P(g_1,g_2,g_3) = \sum_{\v{\jmath},\v{m},\v{n}} {
A^{\v{m}}_{\v{\jmath}}} \sqrt{d_{\v{\jmath}}} \
D^{\v{\jmath}}_{\v{m}\v{n}}(\v{g}) \tj{j}{n},
\end{equation}
where we introduced the Fourier coefficient
\begin{equation}
A^{\v{m}}_{\v{\jmath}}=\sum_{\v{n}}{ \P_{\v{\jmath}}^{\v{m}\v{n}}}{\sqrt{ d_{\v{\jmath}} }}\
  \tj{j}{n},
\end{equation}
and $\tj{j}{n}$ denotes the Wigner $3j$-symbol (see appendix
\ref{app:representations}). The symmetries asked on the fields are
translated on the coefficients :
\begin{eqnarray}
\bar{A}_{j_1j_2j_3}^{m_1m_2m_3} &=& (-1)^{\sum_i (j_i+m_i)}
A_{j_1j_2j_3}^{-m_1-m_2-m_3} \label{eqn:coeffsym1}, \\
A_{j_{\s(1)}j_{\s(2)}j_{\s(3)}}^{m_{\s(1)}m_{\s(2)}m_{\s(3)}} &=&
(-1)^{|\s|} A_{j_1j_2j_3}^{m_1m_2m_3}\label{eqn:coeffsym2}.
\end{eqnarray}

We can now compute the action (\ref{eqn:action}) in terms of the
Fourier coefficients. We insert the Fourier decomposition
(\ref{eqn:fourierdecomp}) in the action, and perform the
integrations using (\ref{eqn:int2mat}). The four $3j$-symbols
appearing in the potential term recombine into a $6j$-symbol. We
finally obtain the action in Fourier modes
\begin{equation}\label{eqn:Boulpot}
S=\frac{1}{2}\sum_{\v{j},\v{n}} |A_{j_1j_2j_3}^{m_1m_2m_3}|^2
+\frac{\l}{4!} \sum_{j_1,...j_6,m_1,...m_6}(-1)^{\sum_i (j_i+m_i)}
A_{j_1j_2j_3}^{-m_1-m_2-m_3} A_{j_3j_5j_4}^{m_3-m_5m_4}
A_{j_4j_2j_6}^{-m_4m_2m_6} A_{j_6j_5j_1}^{-m_6m_5m_1} \sj{j}.
\end{equation}

\begin{figure}
\begin{center}
\includegraphics[height=3cm]{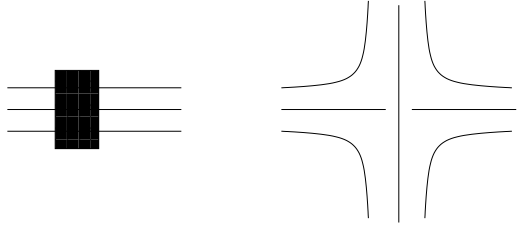}
\end{center}
\caption{Graphical representation of the propagator and potential.
The black square stands for the sum over different permutations of
the three lines, see (\ref{eqn:propagator}). }\label{fig:propvert}
\end{figure}

\begin{figure}[t]
\begin{center}
\includegraphics[height=3cm]{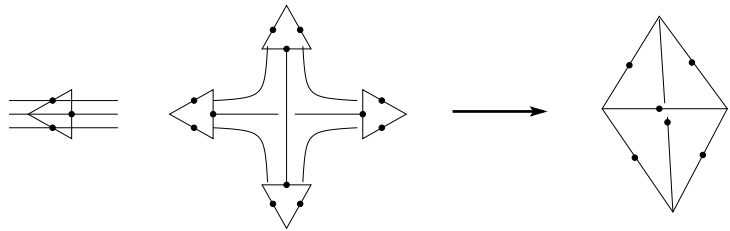}
\end{center}
\caption{Triangulation generated by the Feynman
diagrams}\label{fig:triang}
\end{figure}

We want to compute the partition function and correlation
functions of the theory. They are formally defined in terms of the
path integral \be Z=\int d\P\ e^{-S[\P]}. \ee As usual we compute
this partition function as a perturbative expansion. The Feynman
rules of such a group field model are very similar to those of a
usual QFT. One first construct the propagator  and then  include
the interaction perturbatively. The propagator is given by
\be\label{eqn:propagator} <A^{\v{m}}_{\v{j}}A^{\v{n}}_{\v{j}}> =
{1\over 3!} \sum_{\sigma \in
\Sigma_3}\delta_{\v{j},\sigma(\v{j^\prime})}
\delta_{\v{n},\sigma(\v{m})} \ee where the sum is over permutation
of three elements. The full contribution can be recast into a sum
over Feynman graphs, whose form is dictated by the action. For
each graph we compute its Feynman amplitude by labelling the lines
of the graph  by representations indices; associate a weight to
the lines and the vertices; then sum over representation indices.
In our model the graphical form of the propagator and the vertex
are given in figure \ref{fig:propvert}.
The contributions of the theory at order $n$ are obtained by
constructing graphs with $n$ vertices. We then label each closed
line of the graph by representation indices $(j,m)$, associate to
each vertex the weight given by the $6j$-symbol, then sum over
representation indices. Considering a graph with $n$ vertices and
a symmetry factor $\mathrm{sym}[\G]$, we obtain the amplitude
\begin{equation}
\cA[\G]=\frac{(-\l)^n}{\mathrm{sym}[\G]} \sum_{\{j,m\}} \prod_v
\sj{j}.
\end{equation}
As the weight does not depend on the indices $m$ on the lines, the
sum over each of the $m$ indices is trivial and equal for each
line to the number of terms $\dim j_l=2j_l +1$. Thus one get the
amplitude
\begin{equation}\label{eq:PRweight}
\cA[\G] = \frac{(-\l)^n}{\mathrm{sym}[\G]}\sum_{\{j\}} \prod_l
(2j_l +1) \prod_v \sj{j}.
\end{equation}
By associating a triangle to each propagator
and gluing them together at vertices, one generates tetrahedra (as
seen in figure \ref{fig:triang})
and in general, a Feynman graph of the theory is interpreted as a 3D triangulation.
The state-sum amplitude associated to a triangulation $\D$ is thus
\begin{equation}
\cA[\D] = \frac{(-\l)^n}{\mathrm{sym}[\D]}\sum_{\{j\}} \prod_e
(2j_e +1) \prod_t \sj{j}
\end{equation}
where $e$ and $t$ label the edges and tetrahedron of the
triangulation. This amplitude is exactly the Ponzano-Regge
partition function associated to the triangulation $\D$. Moreover,
all the triangulations can be obtained as Feynman graphs of the
theory, thus the sum over Feynman graphs is indeed the sum over
triangulations of the Ponzano-Regge amplitude.

As in standard quantum field theory, the GFT are in general faced
with two kind of divergences. The first ones are the analog of the
UV/IR divergences, and arise when computing the value of a given
graph by summing over representation indices (momenta). The second
ones are the large order divergences, arising when one try to sum
all the Feynman graphs, in order to get a non-perturbative
expression of a physical quantity. In this paper we will mainly
concentrate on the large order divergences, and we need somehow to
get rid of the UV/IR divergences. For instance, it is well known
that the weight (\ref{eq:PRweight}) is divergent for some
triangulations, the divergences being associated with the vertices
of the triangulation. This is a generalization of the usual fact
that in standard field theory divergences are associated with the
loop of the Feynman graphs (for more details on loop divergences
in group field models see \cite{perez00,perez01}.) As we will see
at the end of this section, one way to avoid these divergences is
to define spin foam models with enhanced convergence properties,
in analogy with the models introduced in \cite{perez00,perez01}.
However the Ponzano-Regge weight generated by the Boulatov model
is not one of these enhanced models, and we feel that a full
understanding of the notion of renormalisability for these
theories should be developed. In this paper we will not dwell
further on this issue and we will ignore the UV problems by
working with a fix regulator and concentrate on the problem of
large order divergences.

The natural regularization to choose in this context, which
respects the symmetry,  is to project out modes of the fields
carrying a spin $j$ bigger than say $N$. So instead of the field
$\P$ we use the projected field
\begin{equation}
\P_N(g_1,g_2,g_3) = \sum_{\v{\jmath}\leq N}
d_{\v{\jmath}}\sum_{\v{m},\v{n}}
  \P_{\v{\jmath}}^{\v{m}\v{n}} D^{\v{\jmath}}_{\v{m}\v{n}}(\v{g}).
\end{equation}
In this case the action expressed in Fourier mode is identical to
(\ref{eqn:Boulpot}), except that the sum over the spins are
restricted to be over $j\leq N$. One can naturally interpret this
projection as arising from a group field theory defined on the non
commutative sphere $S_N^3$ \cite{Ram} (see \cite{madore92} and the
appendix \ref{app:ncsphere} for non commutative 2-sphere $S_N^2$).
This leads to an interpretation of the regularized group field
theory model in terms of a generalized matrix models, since
functions on the non commutative sphere are matrices. Another
natural way to get a regularization is to define this model with
the q-deformation of $\SU(2)$ for $q$ a root of the unity
$q=e^{i\sqrt{\L}}$, where the cut off $N$ in the representations
is related to the cosmological constant $\L$ by $2N+1=\pi
\L^{-1/2}$.

\subsection{Group field model on the sphere and 3D matrix models}
Keeping the action (\ref{eqn:action}), one can define various
models by requiring different kind of invariances on the fields.
For instance by considering a real symmetric field on $\SO(3)^3$
with a $\SO(2)^3$-invariance,
\begin{equation}
\P(g_1h_1,g_2h_2,g_3h_3) = \P(g_1,g_2,g_3)\ \ \forall (h_1,h_2,h_3)\in SO(2)^3
\end{equation}
one obtains a group field model on the coset space
$[\SO(3)/\SO(2)]^3$. To implement the $\SO(2)^3$-invariance, one
fixes an $\SO(2)$ subgroup of $\SO(3)$. In particular, we may
consider $\SO(3)$ in its fundamental representation and choose a
unit vector $v^0$; we then let $H$ be the $\SO(2)$ subgroup which
leaves $v^0$ invariant.  The space of the equivalence classes,
$\SO(3)/\SO(2)$, is diffeomorphic to the 2-sphere $S^2$. Thus a
GFT over the group manifold $[\SO(3)/\SO(2)]^3$ is viewed as
defined over three copies of the 2-sphere $S^2$. As in the case of
the Boulatov model, one has to determine the Fourier transform by
requiring the invariance on the Fourier expansion.
 If the representation is of integer
spin $j$ there is a unique vector invariant under the action of
the $\SO(2)$-subgroup, we denote it $v^0$. The $\SO(2)$-group
averaging of the matrices of representation gives the projector
over this invariant vector
\begin{equation}
\int_{SO(2)} dh\ D^{j}_{mn}(h) = v^0_m v^0_n
\end{equation}
So if we define the Fourier coefficient
\begin{equation}
A_{\v{j}}^{\v{m}} ={\sqrt{d_{\v{\jmath}}}}\
\sum_{\v{n}} \P_{\v{j}}^{\v{m},\v{n}}
v^0_{\v{n}}
\end{equation}
the field is expressed as \be \P(g_1,g_2,g_3) =
\sum_{\v{\jmath},\v{m},\v{n}} A^{\v{\jmath}}_{\v{m}} \sqrt{
d_{\v{\jmath}}}\  D^{\v{\jmath}}_{\v{m}\v{n}}(\v{g})v^0_{\v{n}}.
\ee The reality conditions are translated on the coefficients
\begin{equation}
\bar{A}^{j_1,j_2,j_3}_{m_1m_2m_3}=(-1)^{\sum_i m_i}
A^{j_1,j_2,j_3}_{-m_1-m_2-m_3}
\end{equation}
The action is found to be
\begin{equation}
S=\frac{1}{2}\sum_{j_1,j_2,j_3,m_1,m_2,m_3}
\left|{A_{j_1j_2j_3}^{m_1m_2m_3}}\right|^2
+\frac{\l}{4!}\sum_{j_1...j_6,m_1...m_6} (-1)^{\sum_i m_i}
A_{j_1j_2j_3}^{-m_1-m_2-m_3} A_{j_3j_5j_4}^{m_3-m_5m_4}
A_{j_4j_2j_6}^{-m_4m_2m_6} A_{j_6j_5j_1}^{-m_6m_5m_1}
\end{equation}
The Feynman rules are the same than in the previous case, and the
state sum amplitude associated to a graph is given by
\begin{equation}
(-\l)^n \sum_{\{j, i\}} 1 = (-\l)^n \sum_{\{j\}} \prod_e (2j_e+1).
\end{equation}
As before, this is interpreted as the amplitude associated to a 3D
triangulation. This is a variant of the Ponzano-Regge model, with
$1$ as tetrahedron amplitude, instead of the $6j-$symbol.

This partition function obviously diverges and has to be
regularized. As before, an elegant way to regularize this model is
to consider the original model to be defined on the
non-commutative sphere $S^2_N$ instead of the 2-sphere. The
algebras of function on $S^2_N$ is just the space of $(N+1)\times
(N+1)$ matrices. This space carries naturally a representation of
$\SU(2)$; if we decompose it in terms of spherical harmonics it
can be written as a sum over spin $j$ representation where $j\leq
N$ (see appendix \ref{app:ncsphere}). The amplitude associated to
a graph becomes
\begin{eqnarray}
\cA[\G]&=&\frac{(-\l)^n}{\mathrm{sym}[\G]} \sum_{\{j\leq N\}}
\prod_e (2j_e+1) = \frac{(-\l)^n}{\mathrm{sym}[\G]}
[(N+1)^2]^{(\#\ \textrm{edges})},
\\ &=& \frac{1}{\mathrm{sym}[\G]} (-\l)^{(\#\
\textrm{tetrahedra})} \mu^{(\#\ \textrm{edges})}.
\end{eqnarray}
which is finite (we defined $\u\equiv(N+1)^2$). This amplitude is
the amplitude (\ref{eqn:amplDT}) obtained in 3D dynamical
triangulations. However, we have to notice that to correctly
interpret this GFT as defining a sum over triangulations of the 3D
dynamical triangulation model, we have to set $\l$ negative,
namely $ -\lambda \sim \exp(-\L -{c\over G})$, see
(\ref{eqn:setlambda}).

\subsection{Other group field theories}\label{ssec:otherGFT}
The two previous group field theories were the first ones
introduced in the literature \cite{ambjorn91,boulatov92}. There
has been since then a resurgence of these models when it appeared
that the Barrett-Crane amplitude, which was a candidate for a 4D
state sum model of quantum gravity, could be interpreted as a
Feynman graph amplitude of a group field theory
\cite{freidel00,depietri98}.

Since then, different modifications of the original 4D model have
been proposed first by Perez and Rovelli \cite{perez00} (note also
the generalization in \cite{oriti}). The purpose of these modified
models was to improve the convergence properties of the
 Feynman graph amplitudes. We give here the 3D analogs of these models.

The general group field theory we will consider is  defined on
$\SO(3)^3$ for real symmetric fields, with the action
\begin{eqnarray}
S[\P]&=&\frac{1}{2} \int dg_1dg_2dg_3\ [P_{kin}\P(g_1,g_2,g_3)]^2
\nonumber \\ &+& \frac{\l}{4!} \int  dg_1dg_2dg_3dg_4dg_5dg_6\
[P_{pot}\P(g_1,g_2,g_3)][P_{pot}\P(g_3,g_5,g_4)]
[P_{pot}\P(g_4,g_2,g_6)][P_{pot} \P(g_6,g_5,g_1)],
\end{eqnarray}
where $P_{kin},P_{pot}$ are operators acting on the space of
fields and implementing some invariance property. For instance the
Boulatov model is recovered for $P_{kin}=P_{pot}=P_{SO(3)}$. Where
$P_{SO(3)}\P(g_1,g_2,g_3) = \int_{SO(3)} \P(g_1g,g_2g,g_3g) dg $
is the projector on the space of $SO(3)$ invariant fields. We
denote by $P^R_{SO(2)^3}$ the projector on the sphere, i-e
$P^R_{SO(2)^3}\P(g_1,g_2,g_3) = \int_{SO(2)^3}
\P(g_1h_1,g_2h_2,g_3h_3) dh_1dh_2dh_3$. The Ambjorn model is
recovered for $P_{kin}=P_{pot}=P^R_{SO(2)^3}$. We can consider new
group field theories that gives 3 dimensional analogs of
 the models introduced by Perez and Rovelli, if we use
\begin{eqnarray}
P_{kin} &=& P^R_{SO(3)}, \\ P_{pot} &=& (P^R_{SO(3)}
P^R_{SO(2)^3})^p P_{kin}.
\end{eqnarray}
As the kinetic term is the same than in the Boulatov model, the Fourier coefficient will be
the same. However, applying the projectors in $P_{pot}$ leads to the following state sum amplitude
\begin{equation}\label{eqn:thetastatesum}
\cA[\G]=\sum_{\{j\}} \prod_f (2j_f+1) \prod_v \sj{j}
\T^p(j_1,j_2,j_3)\T^p(j_3,j_5,j_4)\T^p(j_2,j_4,j_6)\T^p(j_1,j_5,j_6)
\end{equation}
where  $\T$ is given by
\begin{equation}
\T(j_1,j_2,j_3)= \left[\tj{j}{n} v^0_{\v{n}} \right]^2= \int dg \
D^{j_1}_{00}(g) D^{j_2}_{00}(g) D^{j_3}_{00}(g).
\end{equation}
One can use the following majoration of $\T$
\begin{eqnarray}
\T(j_1,j_2,j_3)&=&C^{j_1j_2j_3}_{000}C^{j_1j_2j_3}_{000} \nonumber \\
               &\leq& \sum_{m_2 m_3} C^{j_1j_2j_3}_{0m_2m_3}C^{j_1j_2j_3}_{0m_2m_3}
\end{eqnarray}
This last term in the RHS is just $1/d_{j_1}$. Applying this
uniformly for the three spins one gets the bound
\begin{equation}
\T(j_1,j_2,j_3) \leq \frac{1}{(d_{j_1}d_{j_2}d_{j_3})^{1/3}}.
\end{equation}


\section{Non-perturbative resummation of an asymptotic
series}\label{sec:nonpert} In this section we review well known
facts on divergent series in quantum theory. We recall the
physical origin of these divergences, and how to deal with them
from the perturbative and non-perturbative point of view. In
particular, we explain how to obtain a non-perturbative
resummation of a divergent perturbative series. We illustrate
these ideas on a one dimensional $\p^4$ model and show the crucial
role played by the instantons in this framework (see \cite{zinn90}
for a general overview of the large order behavior in quantum
field theory.)

It is known since Dyson \cite{dyson52} that physical quantities in
interacting quantum theory are non analytic functions of the
coupling constant. This is easily related to the fact that for
negative values of the coupling constant, the potential fails to
be bounded from below and the theory becomes unstable. The
consequence is that the perturbative expansion of a physical
quantity in terms of the coupling constant is a divergent series. A
simple example is given by the following action
\begin{equation}\label{eqn:exaction}
S_{\l}(\p)=\frac{1}{2}\p^2+\frac{\l}{4}\p^4,
\end{equation}
which is unbounded from below for negative $\l$. The partition function
\begin{equation}
Z(\l)=\int_{-\infty}^{+\infty} d\p\ e^{-\frac{1}{2}\p^2-
\frac{\l}{4}\p^4} \label{eqn:phi4},
\end{equation}
can be perturbatively evaluated by first expanding the integrand in terms of $\l$, then inverting the integration
and the summation and integrating each term
\begin{equation}
\int_{-\infty}^{+\infty} d\p \ e^{-\frac{1}{2}\p^2}
\p^{4n}=2^{2n+\frac{1}{2}}\G(2n+\frac{1}{2}).
\end{equation}
The exchange of integration and summation is in fact invalid, as a
result the perturbative expansion of $Z(\l)$
\begin{equation}\label{eqn:Zdivseries}
Z(\l)= \sum_n \sqrt{2} (-1)^n \frac{\l^n}{n!} \G(2n+\frac{1}{2}),
\end{equation}
is a divergent series with the n-th term growing like $\cC (-1)^n
4^n n!$.

In general, the asymptotic behavior of perturbative divergent
series can be analyzed by considering classical non-perturbative
objects of the theory called instantons. The instantons are
complex valued solutions of the equations of motion with finite
action. For instance considering the action (\ref{eqn:exaction}),
its equations of motion have two non-trivial complex solutions
given by $\phi=\pm i/\sqrt{\l}$. The value of the action for these
solutions is $S(\phi)=-\frac{1}{4\l}$. These quantities are
expressed using the inverse of the coupling constant, showing the
non-perturbative nature of these objects. Let us see how they play
a role in the large order behavior of perturbative expansions.

\begin{figure}[t]
$$
\includegraphics[width=7cm]{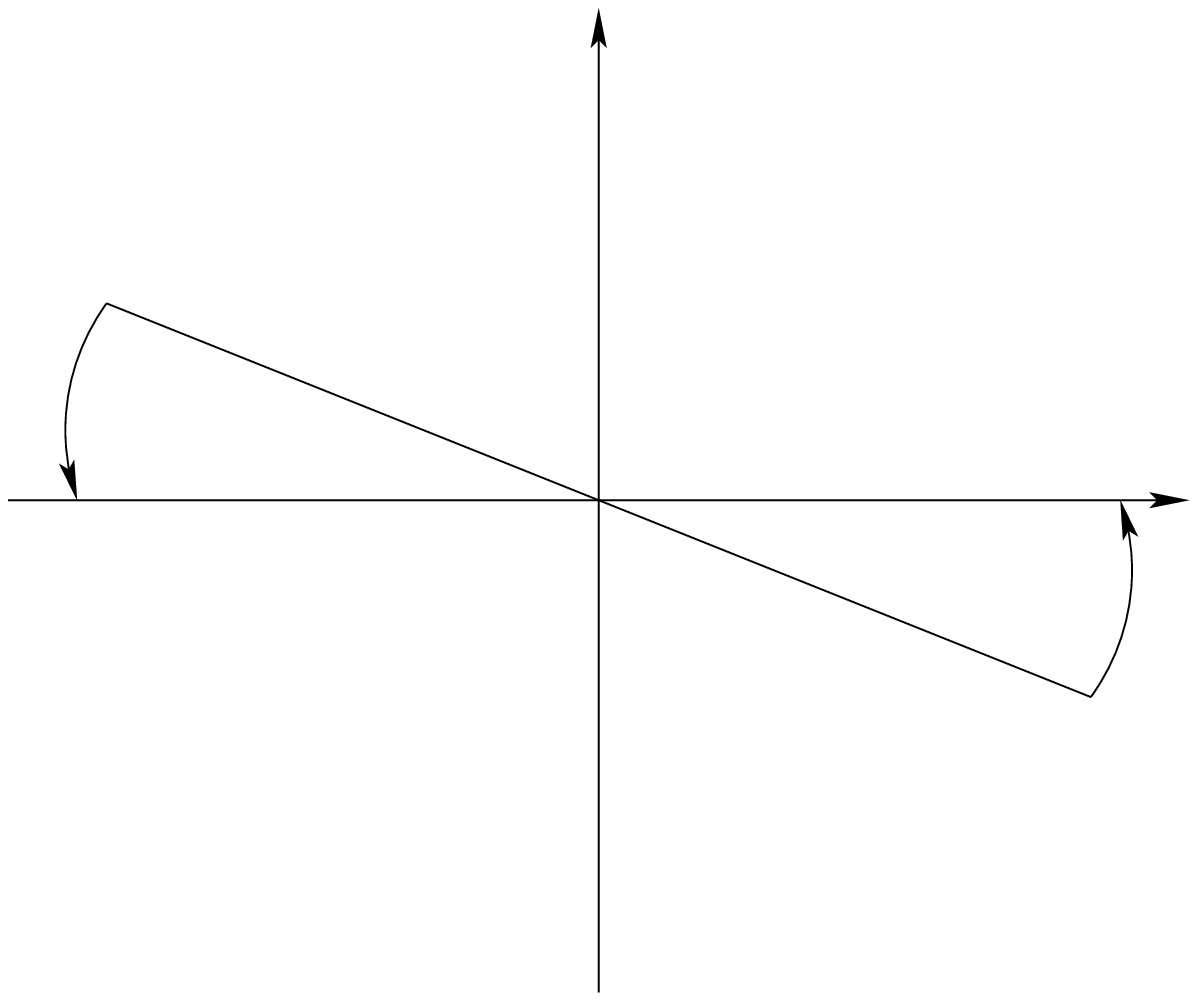}
$$ \caption{Contour for the analytic continuation of
$Z(\l)$}\label{fig:cont1}
\end{figure}

\begin{figure}[t]
\begin{minipage}{0.45\linewidth}
\includegraphics[width=7cm]{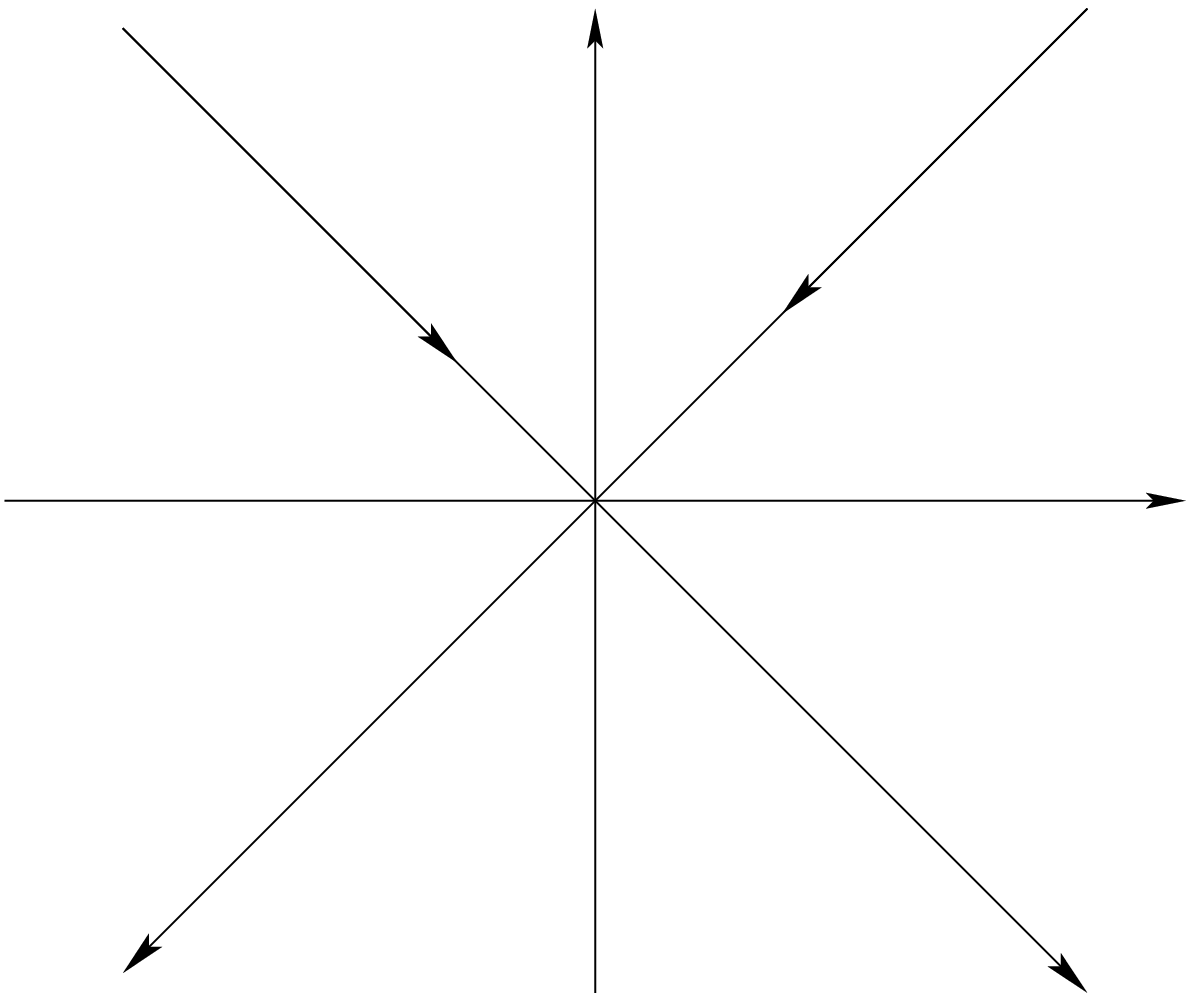}
\end{minipage}
\begin{minipage}{0.45\linewidth}
\includegraphics[width=7cm]{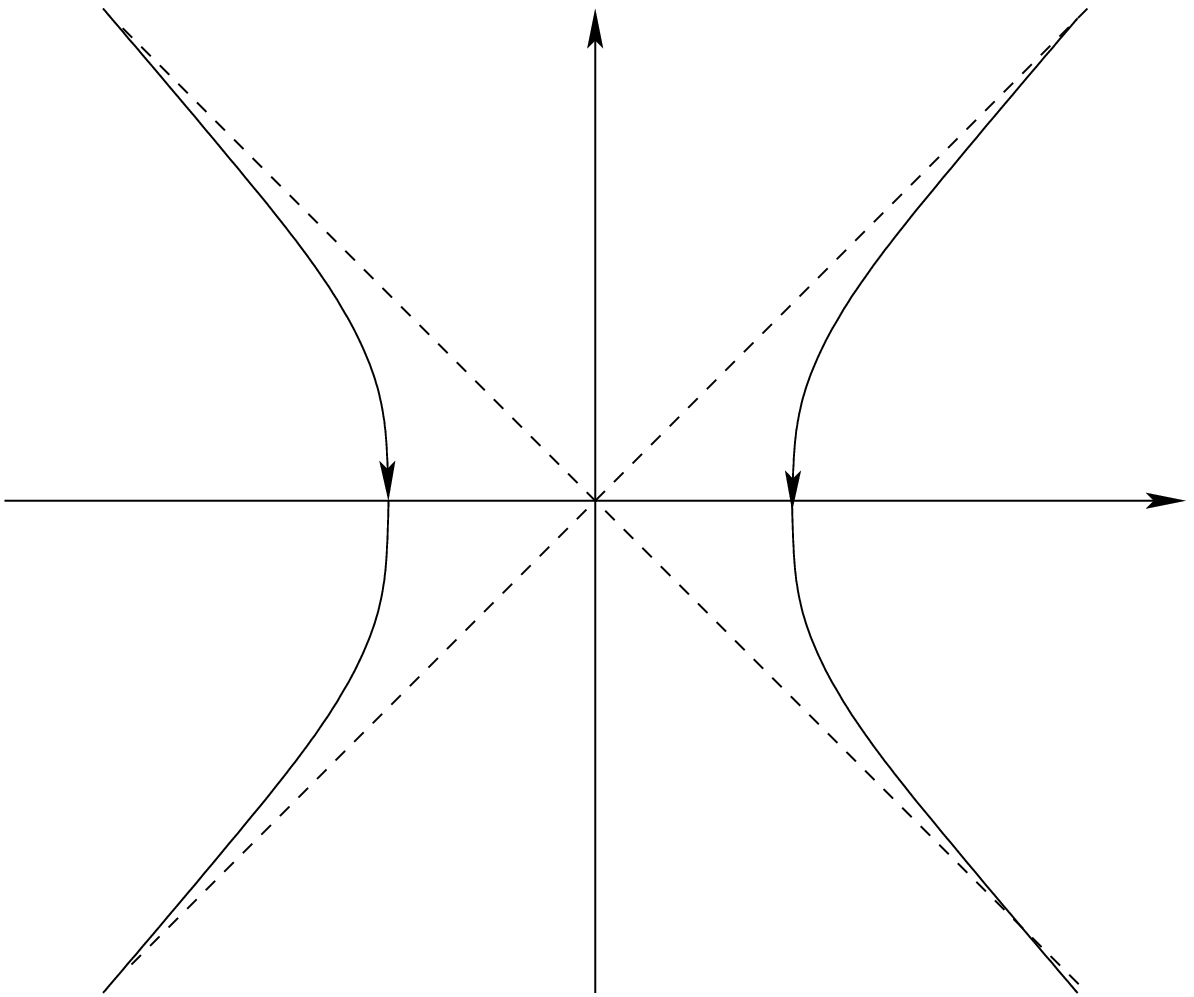}
\end{minipage}
\caption{a) Original contour $C$ for defining $D(-\o)$; b)
Deformed contour $C'$to apply the saddle point method.
}\label{fig:cont2}\label{fig:saddle}
\end{figure}

Consider the previous partition function (\ref{eqn:phi4}). First,
we need to show that it can be analytically continued in the
complex plane, with a cut along the real negative axis. Let $\l=\r
e^{i\t}$ be a complex number. We define $\tilde{Z}(\l)$ by
\begin{equation}
\tilde{Z}(\l) = \int_{e^{-i\t/4}\real} d\phi\  e^{-{1\over 2}
\phi^{2} -{\lambda \over 4}\phi^{4}}
\end{equation}
First $\tilde{Z}(\l)$ is well defined for all $|\theta|<\pi$,
second it is an analytic continuation of $Z(\l)$. This is clear
since for $|\t|<\frac{\pi}{2}$, the integral along
$e^{-i\t/4}\real$ is equal to the integral along $\real$ plus the
circular contours which link these lines at infinity (see figure
\ref{fig:cont1}).
Writing $z=Re^{i\a}$ on these circular contours, the integral
along them is equal to
\begin{equation}
\lim_{R\rightarrow \pm \infty} \int_{-\t/4}^0 d\a\ e^{-\frac{1}{2}
R^2 e^{2i\a}-\frac{\r}{4}e^{i(\t+4\a)} R^4}
\end{equation}
which goes to zero when $R\to \pm \infty$, provided that
$|\t|<\pi/2$. Thus the contour of integration can be shifted from
$e^{-i\t/4}\real$ to $\real$ and the proposed expression defines
the analytic continuation of $Z(\l)$, which can be extended to the
whole complex plane, with a cut on the real negative axis. We can
then define the discontinuity along this cut axis by
\begin{equation}
D(-\o)=\tilde{Z}(e^{i\pi}\o)-\tilde{Z}(e^{-i\pi}\o)
\end{equation}
for $\o>0$. Thus $D$ is given by the integral
\begin{equation}
D(-\o)=\int_{C} d\phi\ e^{-{1\over 2} \phi^{2} + {\o \over
4}\phi^{4}}
\end{equation}
where $C$ is the contour $e^{i\pi/4}\real-e^{-i\pi/4}\real$
pictured in figure \ref{fig:cont2}.
Now $\tilde{Z}$ is related to this discontinuity by the dispersion
relation
\begin{equation}
\tZ(\l)=\frac{1}{2i\pi} \int_0^{+\infty} d\o\ \frac{D(-\o)}{\o+\l}
\end{equation}
We can expand this dispersion relation
\begin{equation}
\tZ(\l)=\sum_n a_n \l^n
\end{equation}
with
\begin{equation}\label{eqn:defcoeff}
a_n=\frac{(-1)^n}{2i\pi} \int_0^{+\infty} d\o\
\frac{D(-\o)}{\o^{n+1}}
\end{equation}
We want to understand the large $n$ behavior of the coefficients
$a_n$ of the series. For large $n$, the integrand in the
definition of these coefficients is sharply peaked around $z=0$.
This shows that the large $n$ behavior of $a_n$ depends on the
value of $D(-\o)$ for small $\o$. We can change the variable of
integration $\psi \equiv \sqrt{\o}\p$.
\begin{equation}
D(-\o)=\int_{C} d\phi\ e^{-{1\over 2} \phi^{2} + {\o \over
4}\phi^{4}} = \int_{C} \frac{d\psi}{\sqrt{\o}} \
e^{\frac{1}{\o}\left(-{1\over 2} \psi^{2} + {1\over
4}\psi^{4}\right)}
\end{equation}
The value of the integral for small $\o$ can be obtained by saddle
point method. The saddle points equation is $-\psi+\psi^3=0$,
whose non-trivial solutions are $\psi=\pm 1$. We see that these
solutions corresponds to instantons solutions, since if we rescale
$\psi$ by $i\sqrt{\o}$, the saddle point equation becomes the
equation of motion of the theory and the solutions becomes the instantons.
The contour $C$ can be deformed
into $C'$ (see figure \ref{fig:saddle}) which contains these
saddle points. Expanding around these points and integrating, one
gets
\begin{equation}
D(-\o) \propto e^{-1/4\o}.
\end{equation}
which is the evaluation of the action for the instantons
solutions. One gets
\begin{equation}\label{eqn:intan}
a_n \propto \cC (-1)^n\int_{0}^{+\infty} d\o\  e^{-\frac{1}{4\o}}
\frac{1}{\o^{n+1}} \propto (-1)^n \left(\frac{1}{4}\right)^{-n} n!
\end{equation}
which is in agreement with the asymptotic given for the exact
computation (\ref{eqn:Zdivseries}). The value $(1/4)^{-n}$ arises
from the evaluation $-1/4z$ of the action on instantons
solutions. This computation shows clearly how the instantons
govern the behavior of the asymptotic series. We will now see how
they are related to the way of extracting perturbative and
non-perturbative information from the divergent expansions.

Even if interacting quantum theories lead to perturbative
expansions which are divergent series one can use these series as
approximation of the non-perturbative quantity by cutting them at
an appropriate order. Consider a divergent series
\begin{equation}
f(z)=\sum_{n=0}^{N-1} a_n z^n +R_N(z)
\end{equation}
with a rest bounded by
\begin{equation}\label{eqn:bndrest}
|R_N(z)|<A S^{-N} N! |z|^N
\end{equation}
(as seen in the previous example, $S$ arise from the integration
(\ref{eqn:intan}) and is related to the evaluation of the action
for first instantons solutions.) One can use the series truncated
at order $N$ as an approximation of the exact non-perturbative
quantity $f(z)$, the difference between them being given by the
rest $R_N(z)$. One can study the variations of $R_N(z)$ with $N$
(at fixed $z$) and show that the rest first decreases as $N$
increases, reaches its minimum around $N \sim S/z$, then increases and
diverge to infinity. This means that the best estimation of $f(z)$
by its truncated asymptotic series is obtained at order $N \sim
S/z$. One can define the accuracy of the expansion by
\begin{equation}
\e(z)=\mathrm{Min}_N\ R_N(z).
\end{equation}
Evaluating the bound (\ref{eqn:bndrest}) for $N \sim S/z$, one
sees that $\e(z)\sim e^{-S/z}$.

One thus know how to extract the best accurate physical
information from a divergent perturbative series. However, one
would like to be able to extract exact finite non-perturbative
information from the divergent perturbative expansions. In order
to get non-perturbative quantities from perturbative expansions,
one has to reconstruct the original function $Z(\l)$ from an
asymptotic series. However, it is easy to see that the issue of
unicity of such reconstruction is not a trivial problem. For
instance, if $f(z)$ as an asymptotic series $\sum a_k z^k$, then
$f(z)+e^{-S/z}$ has the same one. This is not a surprise since one
expects an ambiguity arising from the fact that the accuracy
function is not zero, which means that the perturbative expansion
does not contain all the information necessary to reconstruct the
original function. Thus without additional conditions on $f$, it
remains this ambiguity, there is no way to reconstruct uniquely the
function from the asymptotic series. If one can find conditions on
$f$ to uniquely reconstruct it from its asymptotic series, the
asymptotic series is said to be uniquely Borel summable.

In that case, the unique function is recovered for instance by the
Borel transform procedure. Let us consider an asymptotic series
$\sum a_k z^k$. Assuming that the series
\begin{equation}
B(t)=\sum \frac{a_k}{k!} z^k,
\end{equation}
converges in some circle $|t|<\tau$, and that its analytic continuation admits no singularity
along the positive axis, one can consider the function
\begin{equation}\label{btrans}
f(z)=\int_{0}^{+\infty} e^{-t} B(zt) dt,
\end{equation}
whose perturbative expansion gives the asymptotic series $\sum a_k
z^k$. The possible ambiguity in the reconstruction is manifest in
the Borel transform procedure in the fact that the integral
(\ref{btrans}) can not be unambiguously performed if the Borel
transform $B(t)$ admits poles along the positive axis. In this
case we will have to specify how we go around the poles and
different prescription will differ by terms of the type
$e^{-{S/z}}$.

\begin{figure}
$$
\includegraphics[height=35mm]{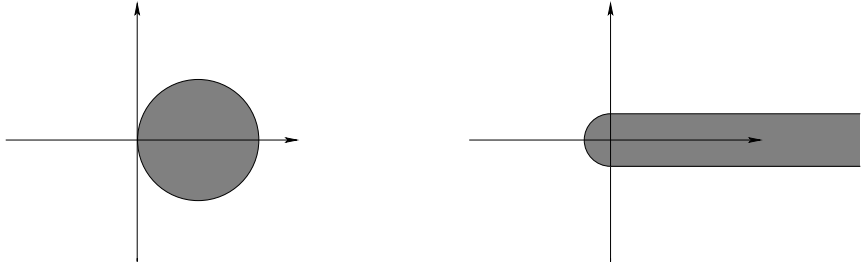}
$$
\caption{Domain $C_R$ of analyticity for $f$, and the
corresponding domain of analyticity for its Borel transform $B_f$,
which contains an open neighbourhood of the real positive
axis.}\label{fig:analyticity}
\end{figure}

The Sokal-Nevalinna's improvement of Watson theorem \cite{sokal80}
gives specific criteria on $f$ in order for its asymptotic series
to be uniquely Borel summable.
\begin{theo}[Sokal-Nevalinna]
Let $f$ be analytic in the circle $C_R=\{z | \Re(z^{-1}) >
R^{-1}\}$ (see figure \ref{fig:analyticity}) and having an
asymptotic expansion
\begin{equation}
f(z)=\sum_{k=0}^{N-1} a_k z^k + R_N(z)
\end{equation}
with
\begin{equation}\label{eqn:theobound}
|R_N(z)| \leq A S^{-N} \G(N+b) |z|^N
\end{equation}
then the asymptotic series is uniquely Borel summable and $f(z)$
can be uniquely reconstructed from it.
\end{theo}
The proof of the theorem can be outlined as follows. Suppose we
have two functions $f_1$ and $f_2$ having the desired asymptotic
expansion and satisfying the hypothesis of the theorem. The
difference between these two functions is given by their rests and
one can bound these difference using (\ref{eqn:theobound}). Our
aim is to prove that this difference is null. Using the minimum on
$N$, it can be bounded by the accuracy of the asymptotic expansion
\begin{equation}
\e(z) \sim e^{-S/|z|}.
\end{equation}
Any such function analytic in the region $C_R$ is in fact null.
For instance the function $e^{-S/z}$ is not analytic in $C_R$ :
let us consider the path $z(t)=t e^{i(\pi/2-t)}$ (which is in
$C_R$ for $t< R$). This path approaches $z=0$ as $t=0$ where as
$\R(1/z(t))= 2\sin t /t \to 2$ so $|\exp^{-S/z(t)}|$ doesn't
approach $0$. From the point of view of the reconstruction via the
Borel transform, the analyticity in $C_R$ ensures the analyticity
of the Borel transform on a strip like region around the real
positive axis (see figure \ref{fig:analyticity}). Thus there is no
ambiguity in the choice of the contour along the real positive
axis.

Coming back to our $\Phi^4$ example, with these criteria one can
prove that the partition function (\ref{eqn:phi4}) has an
asymptotic expansion which is Borel summable. In fact we have
already prove the criteria in our study of the behavior of the
perturbative expansion : we have an analytic continuation in the
cut complex plane which proves the first criterion and an
asymptotic expression of the coefficients of the perturbative
expansion which gives the desired bound for the second criterion.
However, we will give now independent proofs of these criteria
which are simpler and can be adapted to the GFT case.

Concerning the analyticity in the domain $C_R$, if consider
$\l\in\comp$, in the region $\R(\l)>0$, we have
\begin{equation}
|e^{-\frac{\l}{4} \phi^4}|=e^{-\frac{\R(\l)}{4}\phi^4}<1.
\end{equation}
Thus we get the bound
\begin{equation}\label{eqn:boundexp}
\left|\int e^{-\frac{1}{2}\phi^2-\frac{\l}{4} \phi^4}\right|<\int
e^{-\frac{1}{2}\phi^2},
\end{equation}
which ensures the convergence of the integral for $\R(\l)>0$. In
particular, $Z(\l)$ is analytic in a region like the $C_R$ needed
in the first criterion. Concerning the second criterion, we start
from this usual expression for the Taylor rest
\begin{equation}\label{eqn:Taylor}
R_N(\l)=\frac{\l^N}{(N-1)!}\int_0^1 dt\ (1-t)^{N-1} \frac{d^N
Z}{d\l^N}(\l t).
\end{equation}
We use the same bound (\ref{eqn:boundexp}) for $\l\in C_R$ to
obtain
\begin{eqnarray}
\left|\frac{d^N Z}{d\l^N}(\l t)\right| &=&
\left|\int_{-\infty}^{+\infty} d\phi e^{-\frac{1}{2}\phi^2-\frac{\l t}{4}\phi^4}
\frac{\phi^{4N}}{4^N} \right| \\
& \leq & \int_{-\infty}^{+\infty} d\phi e^{-\frac{1}{2}\phi^2}\frac{\phi^{4N}}{4^N} \\
& \leq & \sqrt{2}\ \G(2N+\frac{1}{2})
\end{eqnarray}
This gives for the Taylor rest the bound
\begin{equation}
|R_N(\l)|\leq \frac{\l^N}{N!} \sqrt{2} \G(2N+\frac{1}{2}),
\end{equation}
Due to the multiplicative relation on the $\G$-function
$\G(2x)=\frac{1}{\sqrt{2\pi}}2^{2x-1/2} \G(x)\G(x+1/2)$, one can
find a bound
\begin{equation}
\frac{\G(2N+\frac{1}{2})}{N!}<C A^N \G(N+b),
\end{equation}
which leads a bound of the required form for the rest. The
partition function satisfies the two Sokal criteria, hence its
perturbative expansion is uniquely Borel summable.

\section{Modification of the group field theory model}\label{sec:modif}

\begin{figure}[h]
\begin{center}
\includegraphics[height=35mm]{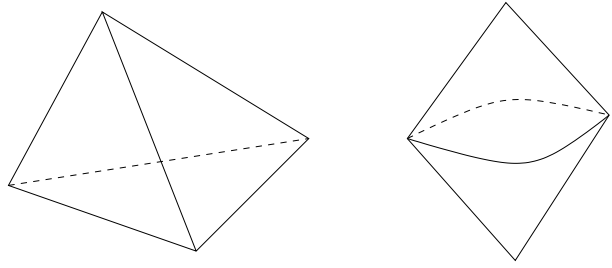}
\end{center}
\caption{The tetrahedron and the pillow : the two objects
constructible from 4 triangles}\label{fig:newtriang}
\end{figure}

In this part we present a new type of group field models, related
to the previous ones, and obtained by changing the potential term
in the action. To determine this potential term, we are guided by
the following ideas. First of all, the previous study of a simple
quantum mechanical case has shown the crucial role played by the
positivity of the potential in the proofs of Borel-summability.
The tetrahedral potential of action (\ref{eqn:action}) has to be
modified since it is not positive. Moreover, it is natural from
renormalisation group arguments to include all the potential terms
of identical degree, which are allowed by the symmetries of the
model (see appendix \ref{app:convgraphs} for a comment on the
divergent graphs which a priori need to be renormalized). From the
point of view of the Feynman graph expansion, this is mirrored in
the fact that the tetrahedron (which appears in the interpretation
of the Feynman graphs as triangulations) is not the only natural
three dimensional object constructible from 4 triangles. One can
construct another object, the "pillow" (see figure
\ref{fig:newtriang}).

Our strategy is thus to include the corresponding interaction term
in the GFT. We consider a general interaction term
\begin{center}
\psfrag{d}{$\d$} \includegraphics[height=35mm]{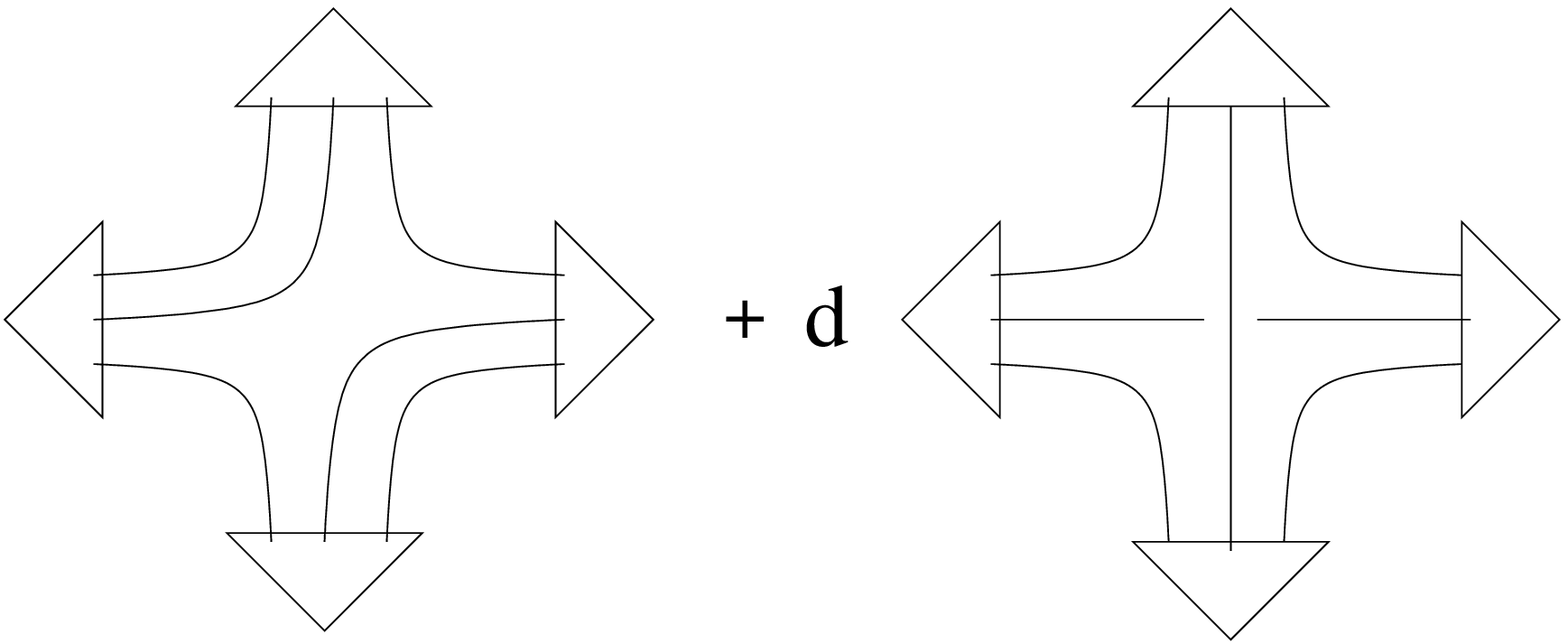}
\end{center}
which corresponds to a total action
\begin{eqnarray}\label{eqn:modifaction}
S[\P]=\frac{1}{2} \int_{G^3} && \P(g_1,g_2,g_3)^2 \nonumber \\+
\frac{\l}{4!}  \int_{G^6}&& \Big[\P(g_1,g_2,g_3)\P(g_3,g_5,g_4)
\P(g_4,g_5,g_6) \P(g_6,g_2,g_1)\nonumber \\ +  && \d\
\P(g_1,g_2,g_3)\P(g_3,g_5,g_4) \P(g_4,g_2,g_6)
\P(g_6,g_5,g_1)\Big]
\end{eqnarray}
Recall now that we are looking for a positive potential. We will
determine the values of $\d$ giving a positive potential. The
strategy is to rewrite the total potential as a scalar product
involving a linear operator whose eigenvalues are under control.
Let us define the object
\begin{equation}
A(g_1,g_2,g_5,g_4) = \int_G dg\ \P(g_1,g_2,g)\P(g,g_5,g_4)
\end{equation}
The pillow term is found to be
\begin{equation}
\int_{G^4} A(g_1,g_2,g_5,g_4) A(g_1,g_2,g_5,g_4) = \langle A,A
\rangle
\end{equation}
while the tetrahedron term can be written as
\begin{equation}
\int_{G^4} A(g_1,g_2,g_5,g_4) [\cT A](g_4,g_2,g_5,g_1) = \langle
A,\cT A \rangle
\end{equation}
where $\cT$ is the just the linear operator acting on $A$
\begin{equation}
\cT : A(g_1,g_2,g_5,g_4) \rightarrow A(g_4,g_2,g_5,g_1).
\end{equation}
This operator $\cT$ is a permutation, thus $\cT^2=Id$ and its
eigenvalues are simply $\pm 1$. The total interaction term we
defined is thus
\begin{equation}
\langle A, [Id+\d\cT]A\rangle
\end{equation}
whose eigenvalues are $1\pm\d$. We obtained a positive potential
for $|\d|\leq 1$. This complete the construction of the modified
potential term.

\begin{figure}[t]
$$
\psfrag{j1}{$j_1$}\psfrag{j2}{$j_2$}\psfrag{j3}{$j_3$}
\psfrag{j4}{$j_4$}\psfrag{j5}{$j_5$}\psfrag{j6}{$j_6$}
\psfrag{k}{$k$}
\includegraphics[height=4cm]{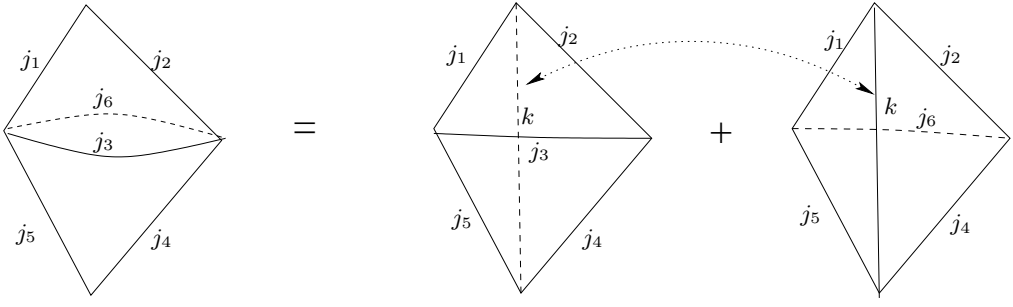}
$$
\caption{A pillow is obtained by the gluing of two tetrahedra
along two common faces, introducing an additional internal edge
$k$}\label{fig:gluetetra}
\end{figure}

At first sight, one could say that such a modification of the
potential term will lead to a drastic modification of the models,
first because we will sum over different kind of triangulations,
second because the weights will be different. We will see that the
modification is not so drastic.  We can first observe that
geometrically, a pillow is no more than the gluing of two
tetrahedra, along two common faces (see figure
\ref{fig:gluetetra}).
This allows us to relate in unique manner each 'generalized'
triangulation (made of pillows and tetrahedra)  to a 'real'
triangulation (made only of tetrahedra) by replacing each pillow
by a pair of tetrahedra glued along two common faces. The total
sum generated by the perturbative expansion can be reorganized by
putting together the weights of all the generalized triangulations
related to a given real triangulation, and summing only over the
real triangulations. The real triangulations containing pairs of
tetrahedra glued along two common faces are called 'irregular',
while the other ones are called 'regular'. We can see that this
reorganization concerns only the irregular triangulations.

\vspace{5mm}

Our results are the following. We consider the modifications of
the Ambjorn and Boulatov GFT given by the action
(\ref{eqn:modifaction}). In both cases the perturbative expansion
of the partition function is uniquely Borel summable. In the case
of the modified Ambjorn model, the perturbative expansion is
written
\begin{equation}\label{DTsum}
Z(\l,\u)=\sum_{\D\in T_{reg}} \frac{1}{\mathrm{Sym}[\Delta]}
\l^{N_3} \u^{N_1}  +\sum_{\D\in T_{irreg}}
\frac{1}{\mathrm{Sym}[\Delta]} \l^{N_3-k} \u^{N_1-k}
\left(\u\l-\frac{1}{2}\right)^k,
\end{equation}
where $\u\equiv(N+1)^2$, $N$ being the cut-off. The first sum is
over regular triangulations, while the second is over irregular
ones. $N_3$ (resp. $N_1$) is the number of tetrahedra (resp.
edges) in $\D$, $k$ denotes the number of pairs of tetrahedra
glued along two common faces in the irregular triangulations. In
the modified Boulatov model, the perturbative expansion can be
written
\begin{equation}\label{PRsum}
Z(\l)=\sum_{\Delta\in T_{reg}} \frac{\l^{n}}{\textrm{Sym}[\Delta]}
Z_{PR}[\Delta]+\sum_{\Delta\in T_{irreg}} \frac{\l^{n-k}
(\l-1/2)^k}{\textrm{Sym}[\Delta]} Z_{PR}[\Delta].
\end{equation}
where $n$ is the number of tetrahedra in the triangulation
$\Delta$, $k$ is the number of pairs of tetrahedra glued along two
common faces in an irregular triangulation and $Z_{PR}[\Delta]$ is
the regularized Ponzano-Regge amplitude, for the triangulation
$\D$. These two formulas have been written for the parameter
$\d=-1$ in the potential (see \ref{eqn:modifaction}), but we will
give a general expression valid for $|\d|<1$ in the proof in the
next section.

Several remarks can be made. First, despite the modification made
on the GFT models, we still obtain a perturbative expansion which
can be written as a sum over triangulations. Moreover, for all
regular triangulations which do not contain any pair of tetrahedra
glued along two common faces, these weights are the same than the
original models. The weight are modified for the irregular
triangulations and can even be cancelled by an appropriate choice
of the coupling constant, namely $\l=1/2$ in the Boulatov model or
$\u\l=1/2$ in the dynamical triangulation model. In addition, our
modification lead to another surprising consequence. The
modification of the potential we choose leads to the fact that the
factor depending on the coupling constant is now $\l^n$ instead of
$(-\l)^n$, as it was the case for the originals models. Thus the
relevant regime is \textit{for positive values of $\l$}, namely
the physical region of convergence, where the GFT is well-defined.
Our choice of positive potential allows us not only to prove the
Borel summability of the model, but also to have a well-defined
GFT for the values of the coupling constant which are physically
relevant.

\section{Proof of the Borel-summability of the perturbative
expansion}\label{sec:proof} In this part we prove the main result,
namely the unique Borel summability of the two modified models, and the
fact that the resulting perturbative expansions are rewritten as
slight modifications of the original models. The
proofs of Borel summability are closed to those given in the
simple quantum mechanical $\Phi^4$ case at the end of section
\ref{sec:nonpert}.

\subsection{Sokal criteria for the modified models}
The analyticity of the partition function for $\R(\l)\geq 0$
follows directly from the positivity of the potential, using the
same bound than in the $\Phi^4$ case, eq.(\ref{eqn:boundexp}).
This proves the first criterion.

We consider now the Fourier expansion of these models. For the 3D
matrix model, we find
\begin{eqnarray} S=\frac{1}{2}\sum_{\vj,\vm} |A_{\vj}^{\vm}|^2
+\frac{\l}{4!}\sum_{j_1...j_6,m_1...m_6} (-1)^{\sum_i m_i} \Big(&&
A_{j_1j_2j_3}^{-m_1-m_2-m_3} A_{j_3j_5j_4}^{m_3-m_5m_4}
A_{j_4j_5j_6}^{-m_4m_5m_6} A_{j_6j_2j_1}^{-m_6m_2m_1}\nonumber \\+
\d && A_{j_1j_2j_3}^{-m_1-m_2-m_3} A_{j_3j_5j_4}^{m_3-m_5m_4}
A_{j_4j_2j_6}^{-m_4m_2m_6} A_{j_6j_5j_1}^{-m_6m_5m_1}\Big)
\label{eqn:fourierexp}
\end{eqnarray}
For the Boulatov model, we find
\begin{eqnarray}
S = \frac{1}{2} \sum_{\v{j},\v{n}}&&
\left|A^{\v{m}}_{\v{\jmath}}\right|^2+ \frac{\l}{4!}
\sum_{j_1,...j_6,m_1,...m_6}(-1)^{\sum_i (j_i+m_i)}\nonumber \\
\Big(&&A_{j_1j_2j_3}^{-m_1-m_2-m_3} A_{j_3j_5j_4}^{m_3-m_5m_4}
A_{j_4j_5j_6}^{-m_4m_5m_6} A_{j_6j_2j_1}^{-m_6m_2m_1}
\t_{j_1j_2j_3}\t_{j_1j_2j_6}\t_{j_4j_5j_3}\t_{j_4j_5j_6}\frac{\d_{j_3,j_6}}{2j_3+1}
\nonumber \\ +\d && A_{j_1j_2j_3}^{-m_1-m_2-m_3}
A_{j_3j_5j_4}^{m_3-m_5m_4} A_{j_4j_2j_6}^{-m_4m_2m_6}
A_{j_6j_5j_1}^{-m_6m_5m_1} \sj{j}\Big)
\end{eqnarray}
where $\t_{abc}=1$ if $a,b,c$ satisfies the triangular
inequalities, $0$ if not.

In order to clarify the notations and give a uniform proof for the
two cases, we separate the coefficients into real and imaginary
parts. All the real coefficients we obtained are not independent
due to the relations (\ref{eqn:coeffsym1},\ref{eqn:coeffsym2}). We
denote by $\{A_I, I=1...d\}$ a set of independent coefficients ($d$
is related to the cutoff $N$ and behaves roughly as $N^6$, as it
stands for the six indices $(j_1,j_2,j_3,m_1,m_2,m_3)$, up to
symmetry relations). The action expressed in Fourier modes can be
rewritten as
\begin{equation}
S=\frac{1}{2}\sum_{I} A_{I}^2+\frac{\l}{4!} \sum_{IJKL} A_I A_J
M^{IJKL} A_K A_L
\end{equation}
with $M^{IJKL}$ a four indices matrix representing the potential,
whose positivity is translated into
\begin{equation}
\sum_{IJKL} A_I A_J M^{IJKL} A_K A_L \geq 0\ \ \forall A_I
\end{equation}
The partition function is
\begin{equation}\label{eqn:partfourier}\label{eqn:defz}
Z(\l)=\int_{\real^d} \left[\prod_I dA_I
\right]\exp\left\{-\frac{1}{2}\sum_I A_I^2 -\frac{\l}{4!}
\sum_{IJKL} A_i A_J M^{IJKL} A_K A_L \right\}
\end{equation}

We want to consider the asymptotic expansion of $Z(\l)$
\begin{equation}
Z(\l)=\sum_{p=0}^{n-1} a_p \l^p + R_n(\l)
\end{equation}
We use the expression (\ref{eqn:Taylor}) for the Taylor rest.
Consider the expression for $\frac{d^nZ}{d\l^n}$ obtained by
deriving $n$ times the original expression (\ref{eqn:partfourier})
\begin{equation}
\frac{d^nZ}{d\l^n}(\l t)=\int_{\real^d} [\prod_I dA_I]\ \exp\left[
-\frac{1}{2}\sum_I A_I^2 -\frac{\l t}{4!} \sum_{IJKL} A_I A_J
M^{IJKL} A_K A_L \right] \left(-\frac{1}{4!} \sum_{IJKL} A_I A_J
M^{IJKL} A_K A_L\right)^n
\end{equation}
One look for a bound of the module of the integrand. The real part
of the potential is positive for $\l\in C_R$, thus
\begin{equation}
\left|\exp\left[ -\frac{1}{2}\sum_I A_I^2 -\frac{\l t}{4!}
\sum_{IJKL} A_I A_J M^{IJKL} A_K A_L \right]\right|<\exp\left[
-\frac{1}{2}\sum_I A_I^2\right]
\end{equation}
Thus one obtains the following bound
\begin{equation}
\left|\frac{d^nZ}{d\l^n}(\l t)\right| \leq \int_{\real^d} [\prod_I
dA_I] \exp\left[ -\frac{1}{2}\sum_I  A_I^2\right]
\left|\frac{1}{4!} \sum_{IJKL} A_I A_J M^{IJKL} A_K A_L\right|^n
\end{equation}
Let us denote
\begin{equation}
S=\mathrm{Sup}_{IJKL}\ |M^{IJKL}|
\end{equation}
One gets the bound
\begin{equation}\label{eqn:bnd}
\left|\frac{d^nZ}{d\l^n}(\l t)\right| \leq \frac{S^n}{(4!)^n}
\int_{\real^d} [\prod_I dA_I] \exp\left[ -\frac{1}{2}\sum_I
A_I^2\right] \left(\sum_{IJKL} |A_I A_J A_K A_L|\right)^n
\end{equation}
The RHS integral is symmetric over the $A_I$, thus one can write
it as $d!$ times the integral over the domain $A_1>A_2>...>A_{d}$
(recall that $d$ denotes the total number of Fourier modes and
depends polynomially on the cut-off $N$). In that case,
\begin{equation}
\sum_{IJKL} |A_I A_J A_K A_L|<d^4 |A_1|^4
\end{equation}
Then we can decouple the integrals over the $A_I$ on LHS of
(\ref{eqn:bnd}) and extend the integral for $I\neq 1$ in order to
get $d-1$ gaussian integrals. The integral over $A_1$ gives a
$\G-$function
\begin{equation}
\left|\frac{d^nZ}{d\l^n}(\l t)\right| \leq d! \left(\frac{S\
d^4}{4!}\right)^n \left(\int_\real dA_1\  e^{-\frac{1}{2}A_1^2}
A_1^{4n}\right) \left(\int_\real dA\ e^{-\frac{1}{2}
A^2}\right)^{d-1}
\end{equation}
and
\begin{equation}
\left|\frac{d^nZ}{d\l^n}(\l t)\right| \leq \left(\sqrt{2}\ d!\
\pi^{\frac{d-1}{2}}\right) \left(\frac{S\
d^4}{3!}\right)^n\G(2n+\frac{1}{2})
\end{equation}
Coming back to the expression of the Taylor rest
(\ref{eqn:Taylor}), one gets a bound
\begin{equation}
|R_n(\l)| \leq \cC\ \left(\frac{S\ d^4}{3!}\right)^n
\frac{\G(2n+\frac{1}{2})}{n!}|\l|^n
\end{equation}
which is of the required form, with constants depending on $d$,
hence on the cut-off $N$. The perturbative expansion of our model
satisfies the two Sokal criteria and thus is uniquely Borel
summable.

\subsection{Reorganization of the perturbative expansion}

We now analyze the consequences of the introduction of the pillow
interaction term on the expression of the sum over triangulations.
The perturbative expansions of modified GFT models give us a sum
over triangulations made with pillows and tetrahedra. However, we
have seen that geometrically a pillow is obtained by gluing
together two tetrahedra along two common faces.

In the matrix model case, the amplitude associated to a pillow is
\begin{equation}
A_{pillow}=-\l
\end{equation}
while the amplitude for two tetrahedra glued along two common
faces (see figure \ref{fig:gluetetra}) is
\begin{equation}
A_{2tetra}=(-\d\l)^2 \sum_{k\leq N} (2k+1) = (\d\l)^2 (N+1)^2
\end{equation}
Thus, if we compare the total amplitude for a (generalized)
triangulation $\tilde\D$ containing $k$ pillows to the related
real triangulation $\D$ obtained by replacing all the pillows by
pairs of tetrahedra, we obtain (recall $\u=(N+1)^2$)
\begin{equation}
Z[\tilde\D]=\left(-\frac{1}{2}\frac{1}{\l \d^2 \u}\right)^k Z[\D]
\end{equation}
The one-half factor comes from the fact that the symmetry factor
for two tetrahedra is twice the symmetry factor for one pillow.
Consider a real triangulation $\Delta$ made only with $N_3$
tetrahedra and $N_1$ edges, and such that $2k$ of its tetrahedra
are glued by pair along two of their common faces. There are
several generalized triangulations related to it : $C_k^1$
(binomial coefficients) triangulations with one pair of such
tetrahedra replaced by a pillow, $C_k^2$ triangulations with two
such pairs replaced by pillow etc...Counting together all their
weights we obtain
\begin{equation}
Z[\D]=(-\d\l)^{N_3}\Big(1+(-2\d^2\l\u)^{-1}C_k^1+(-2\d^2\l\u)^{-2}C_k^2+...+(-2\d^2\l\u)^{-k}C_k^k\Big)\u^{N_1}=(-\d\l)^{N_3}
(1-(2\d^2\l\u)^{-1})^k \u^{N_1}
\end{equation}
This shows that one can reorganize our sum as a sum over real
triangulations of the amplitude $(-\d\l)^{N_3}\u^{N_1}$, except
the fact that an irregular triangulation containing $k$ pairs of
tetrahedra glued together along two common faces has a weight
modified by a factor $(1-(2\d^2\l\u)^{-1})^k$. Which gives
\begin{equation}
Z(\l,\u)=\sum_\D  \frac{1}{\mathrm{Sym}[\Delta]} (-\d\l)^{N_3-k}
\u^{N_1-k} \left(-\d\u\l+\frac{1}{2\d}\right)^k.
\end{equation}
Recall that the positivity of the potential is valid for $|\d|\leq
1$. For instance for $\d=-1$, we recover the result (\ref{DTsum}) announced in
section \ref{sec:modif}.

The same reorganization can be performed on the perturbative
expansion of the modified Boulatov model. The amplitude associated
to a pillow is
\begin{equation}
A_{pillow}=-\l \frac{\d_{j_3,j_6}}{2j_3+1}
\end{equation}
The amplitude associated to the gluing of two tetrahedra along two
common faces (see figure \ref{fig:gluetetra}) is
\begin{equation}
A_{2tetra}=(-\d\l)^2 \sum_k (2j_k+1) \left\{\begin{array}{ccc} j_1 & j_2 & k \\
j_4 & j_5 & j_3
\end{array}\right\}
\left\{\begin{array}{ccc} j_1 & j_2 & k \\
j_4 & j_5 & j_6
\end{array}\right\}=(\d\l)^2\frac{\d_{j_3,j_6}}{2j_3+1}
\end{equation}
due to the orthogonality relation. As before, this means that a
triangulation containing a pillow has an amplitude which is
$(1-2\d^2\l)^{-1}$ times the amplitude of the same triangulation,
with the pillow replaced by two tetrahedra glued along two common
faces. We reorganize the sum in the same way than before and
obtain
\begin{equation}
Z(\l)=\sum_\D \frac{1}{\mathrm{Sym[\D]}}(-\d\l)^n
(1-\frac{1}{2\d^2\l})^k Z_{PR}[\Delta] = \sum_\D
\frac{1}{\mathrm{Sym[\D]}} (-\d\l)^{n-k} (-\d\l+\frac{1}{2\d})^k
Z_{PR}[\D].
\end{equation}
The result (\ref{PRsum}) given in section \ref{sec:modif} is recovered for
$\d=-1$.

\section{Conclusion}\label{sec:conc}
In this paper we have studied the possibility of defining the sum
over topologies of three dimensional gravity in the context of
dynamical triangulation or spin foam models. We have provided an
example of a group field theory model which is uniquely Borel
summable in the physical regime of the theory. In order to do so
we have used the fact that three dimensional triangulations  are
generated by a group field theory with a quartic potential and we
have shown that this potential can be naturally modified in order
to be positive. Therefore, the message we wanted to convey here
is that it is possible to resum the sum over triangulation  if
one accepts to take for serious the non perturbative information
provided to us  by  the GFT, namely the instantonic solutions.
One can still question what is the physical content of this non
perturbative information. This is a direction of investigation we
are following. Finally, 4 dimensional triangulations are generated
by quintic potentials and we may think that there is no natural
modification of this potential leading to a uniquely resummable
theory.  This is an issue that deserves to be studied and could
mean that we need more non-perturbative input in this case in
order to give a meaning to the resummed series.

\vspace{1cm}

\noindent\textbf{Acknowledgments :} We would like to thank J.
Ambjorn for a discussion. D. L. is supported by a MENRT grant and
Eurodoc program from R\'egion Rh\^one-Alpes. L. F. is supported by
CNRS and an ACI-Blanche grant.

\appendix
\section{Representations of $\SO(3)$ and Racah-Wigner symbols}\label{app:representations}

The irreducible unitary inequivalent representations of $\SU(2)$
are labelled by a half-integer $j$. The representations of
$\SO(3)$ are just the integer spin representation of $\SU(2)$. The
space of representation is denoted $V_j$ and has dimension
$\dim_{j} = 2j+1$. All the representations are self-dual
$V_j^*=V_j$. The matrix of representations are denoted
$D^j_{mn}(g)$. We have the relation
\begin{equation}
\bar{D}^j_{mn}(g)= (-1)^{m-n} D^j_{-m-n}(g)
\end{equation}
Given three representations $V_{j_1},V_{j_2},V_{j_3}$, it exists
only one (up to normalization) intertwiner $\imath : V_{j_1}
\otimes V_{j_2} \otimes V_{j_3} \to \comp$. One can express it by
using the Wigner $3j$-symbol whose normalization is
\begin{equation}
\sum_{\vm} \tj{j}{m} \tj{j}{m} = 1
\end{equation}
The Racah-Wigner $6j$ symbol is defined by
\begin{equation}
\sj{j}=\sum_{m_i} (-1)^{j_4+j_5+j_6+m_4+m_5+m_6} C^{j_1  j_2
j_3}_{m_1  m_2  m_3} C^{j_5  j_6 j_1}_{m_5  -m_6  m_1} C^{j_6 j_4
j_2}_{m_6  -m_4  m_2} C^{j_4  j_5 j_3}_{m_4  -m_5  m_3}
\end{equation}
It satisfies the  orthogonality relation
\begin{equation}\label{eqn:orth6j}
\sum_{j_6} (2j_3+1) (2j_6+1) \sj{j} \left\{\begin{array}{ccc} j_1
& j_2 & j'_3 \\ j_4 & j_5 & j_6
\end{array}\right\}
= \d_{j_3 j'_3} \t_{j_1 j_2 j_3} \t_{j_3 j_4 j_5}
\end{equation}
where $\t_{abc}=1$ if $a,b,c$ satisfy the triangular inequalities,
0 in the contrary case. The fact that the integral of a matrix
representation gives a projection on the space of invariant
vectors leads to the following relations (using the normalized
Haar measure)
\begin{equation}\label{eqn:int2mat}
\int_{\SU(2)} dg D^j_{mn}(g) \bar{D}^{j'}_{m'n'}(g) =
\frac{1}{d_j} \d^{jj'}\d_{mm'}\d_{nn'},
\end{equation}
and
\begin{equation}\label{eqn:int3mat}
\int dg D^{j_1}_{m_1n_1}(g) {D}^{j_2}_{m_2n_2}(g)
D^{j_3}_{m_3n_3}(g)=\tj{j}{m}\tj{j}{n}.
\end{equation}

\section{The non-commutative 2-sphere} \label{app:ncsphere}
A non-commutative space $X$ is a space which is \textit{defined}
by specifying the (non-commutative) algebra of square integrable
functions on it $L^2(X)$. In this part we recall some facts about
the construction of the non-commutative sphere. The sphere $S^2$
can be described by the algebra $L^2(S^2)$. A basis of this
algebra is given by polynomes of $(x_1,x_2,x_3)$ such that
\begin{equation}
x_1^2+x_2^2+x_3^2=1 \label{eqn:sumsquare}
\end{equation}
Another possible basis is given the spherical harmonics $Y^{lm}$ and arises from the
decomposition
\begin{equation}
L^2(S^2)=\bigoplus_{l=1}^{+\infty} V_l.
\end{equation}
where $V_l$ is the spin $j$ representation of $\SO(3)$. Let us
construct the non-commutative sphere $S^2_N$ (for a deformation
parameter $1/N$) by giving its algebra of functions, obtained by
starting from the first basis described for the sphere. Let us
consider $V_{\frac{N}{2}}$, the spin $N/2$ representation of the
algebra $\su(2)$; $D^{N/2}$ are the matrices of representation. Let
us define
\begin{equation}
X_i=\frac{1}{N} D^{N/2}(J_i)
\end{equation}
where the $J_i$ are the generators of $\su(2)$. $X_i$ is in
$V_{\frac{N}{2}}$. Due to the Lie algebra relations of $\su(2)$,
one gets
\begin{eqnarray}
\sum_i X_i X^i &=& 1 \\ \left[X_i,X_j\right]&=&\frac{1}{N}
\e_{ijk} X_k
\end{eqnarray}
The first relation is the analogue of the relation (eq. \ref{eqn:sumsquare}), and the second one
says that the $X_i$ are commuting in the large $N$ limit. This facts lead us to see the $X_i$ as
non-commuting spherical coordinates, and to define the algebra of functions on the non-commutative
sphere as the algebra of endomorphism of $V_{\frac{N}{2}}$
\begin{equation}
L^2(S^2_N) = End(V_{\frac{N}{2}}) \sim V_{\frac{N}{2}} \otimes V^*_{\frac{N}{2}}
\end{equation}
The right-hand side can be written as
\begin{equation}
\bigoplus_{l=0}^N V_l,
\end{equation}
which shows that the parameter $N$ is nothing more than the
cut-off in the decomposition of the functions over the sphere into
spherical harmonics. As the endomorphisms of $V_{\frac{N}{2}}$ are
the non-commutative counterpart of the polynomials over the
sphere, a natural question is to ask what are the non-commutative
counterpart of the spherical harmonics basis. Indeed, each
spherical harmonic $Y^{l,m}$ for $l \leq N$ can be represented as
an endomorphism of $V_{\frac{N}{2}}$ by
\begin{equation}
Y^{l,m} \longrightarrow [\hat{Y}^{lm}]_{ij}=C^{l\ \frac{N}{2}\
\frac{N}{2}}_{m\ i\ j}.
\end{equation}
\section{Note on the convergence of graphs}\label{app:convgraphs}

If we consider the amplitude associated to a graph $\G$ in the
group field theory obtained by using $p$ times the projection (see
subsection \ref{ssec:otherGFT}) , we get
\begin{equation}
\cA[\G]=\sum_{\{j\}} \prod_f d_{j_f} \prod_e \T^p(j_1,j_2,j_3)
\prod_v \sj{j}\label{eqn:amplconv}
\end{equation}
We have proven for the $\T$ the following bound
\begin{equation}
\T(j_1,j_2,j_3) \leq \frac{1}{(d_{j_1}d_{j_2}d_{j_3})^{1/3}}.
\end{equation}
We consider also the bound on the $6j$-symbol
\begin{equation}
\left|\sj{j}\right|\leq\frac{1}{(\prod_i^6 d_{j_i})^{1/6}}
\end{equation}
obtained from the orthogonality relation (\ref{eqn:orth6j}).

The amplitude (\ref{eqn:amplconv}) can be bounded
\begin{equation}
\cA[\G] \leq \sum_{\{j_f\}} \prod_f
d_{j_f}^{1-\frac{p}{3}n_e(f)-\frac{1}{6}n_v(f)}
\end{equation}
where $n_e(f)$ and $n_v(f)$ are the number of edges and vertices
lying in the face $f$. As these numbers are equal, we obtain the
bound
\begin{equation}
\cA[\G] \leq \prod_f \left(\sum_{j_f}
d_{j_f}^{1-\frac{2p+1}{6}n_v(f)}\right).
\end{equation}
One can see that there is no divergence associated to the sum over
spins of face $f$ if the number of vertices lying in this face is
such that
\begin{equation}
n_v(f) > \frac{12}{2p+1}.
\end{equation}

For instance in the case of $p=2$, we obtain that the only graphs
which are not proved to be convergent are for the faces such that
$n_v(f)=1,2$. The dual faces bounded by only one or two dual
vertices correspond to edges bounded by only one or two tetrahedra
in the triangulation. This observation justify that only these two
types of graphs need to be renormalized. However, our bounds are
not optimal and numerical simulations suggest that all the graphs
actually converge for $p=2$, while for $p=1$, the graphs with
$n_f(v)=1,2$ diverges.


\end{document}